\documentclass[prl,notitlepage,twocolumn]{revtex4}

\usepackage{amsmath}
\usepackage{amssymb}
\usepackage{bm}
\usepackage{graphicx}
\usepackage{times}

\usepackage{color}

\begin{document}

\title{Disorder-immune photonics based on Mie-resonant dielectric metamaterials}

\author{Changxu Liu$^{1,2}$}
\author{Mikhail V. Rybin$^{3,4}$}
\author{Peng Mao$^{1,5}$}
\author{Shuang Zhang$^1$}
\author{Yuri Kivshar$^{3,6}$}

\affiliation{$^1$School of Physics and Astronomy, University of Birmingham, Birmingham B15 2TT, UK}
\affiliation{$^2$Chair in Hybrid Nanosystems, Nanoinstitute Munich, Faculty of Physics, Ludwig Maximilians University of Munich, 80539 Munich, Germany}
\affiliation{$^3$ITMO University, St Petersburg 197101, Russia}
\affiliation{$^4$Ioffe Institute, St Petersburg 194021, Russia}
\affiliation{$^5$College of Electronic and Optical Engineering and College of Microelectronics, Nanjing University
of Posts and Telecommunications, Nanjing 210023, China}
\affiliation{$^6$Nonlinear Physics Centre, Australian National University, Canberra, ACT 2601, Australia}

\begin{abstract}
When the feature size of photonic structures becomes comparable or even smaller than the wavelength of light, the fabrication imperfections inevitably introduce disorder that may eliminate many functionalities of subwavelength photonic devices. Here we suggest a novel concept to achieve a robust bandgap which can endure disorder beyond $30\%$ as a result of the transition from photonic crystals to Mie-resonant metamaterials. By utilizing Mie-resonant metamaterials with high refractive index, we demonstrate photonic waveguides and cavities with strong robustness to position disorder, thus providing a novel approach to the bandgap-based nanophotonic devices with new properties and functionalities.
\end{abstract}


\maketitle

The idea of manipulating the electromagnetic waves with subwavelength structures originates from the 19-th century, when Heinrich Hertz managed to control meter-long radio waves through wire-grid polarizer with centimeter spacings~\cite{trager2012springer}.
As the rapid advancement of the nanotechnology with fabrication resolution down to micrometer or even nanometers, a plethora of subwavelength systems with structure-induced optical properties are achieved, ranging from photonic crystals to metamaterials \cite{cheben2018subwavelength}.
Among them, a photonic crystal (PhC) is a periodic optical structure that has attracted considerable interest for its ability to confine, manipulate, and guide light \cite{joannopoulos2011photonic}. Spatial periodicity of the dielectric function  is essential to obtain {\it a photonic bandgap} where the propagation for photons within a certain frequency gap is forbidden, providing unique features for a variety of applications ranging from lasers \cite{hirose2014watt, wu2015monolayer}, all-optical memories \cite{kuramochi2014large} to sensing \cite{fenzl2014photonic} and emission control \cite{noda2007spontaneous}.

To achieve unparalleled functionalities with the subwavelength structures, a stringent requirement for the fabrication accuracy is required.
As a result, the impact of disorder on such photonic structures has extensively been studied, both numerically and experimentally  ~\cite{astratov2002interplay, koenderink2003light, zharov2005suppression, hughes2005extrinsic, rockstuhl2006correlation, gollub2007characterizing, asatryan2007suppression, topolancik2007experimental, toninelli2008exceptional,engelen2008two, rockstuhl2009suppression, garcia2010density,sapienza2010cavity, garcia2011photonic,muskens2011broadband, savona2011electromagnetic, li2001photonic, li2000fragility, bayindir2001photonic, kaliteevski2002disorder,  yamilov2004highest,rengarajan2005effect, kaliteevski2006stability, lavrinenko2009influence, dorado2009modeling,alu2010effect, mogilevtsev2011light, maguid2017disorder, jang2018wavefront,poddubny2012fano}.
When the disorder is small enough (up to a few percent of the lattice constant) and can be treated as a perturbation, the interaction between the order and disorder gives rise to interesting optical transport phenomena involving multiple light scattering, diffusion and localization of light~\cite{topolancik2007experimental, toninelli2008exceptional,engelen2008two, garcia2010density,sapienza2010cavity, garcia2011photonic,muskens2011broadband, savona2011electromagnetic}.
As disorder is increased further, the photonic bandgap is destroyed, owing to the adverse effect to the Bragg reflection \cite{li2001photonic, li2000fragility, bayindir2001photonic, kaliteevski2002disorder,  yamilov2004highest, rengarajan2005effect, kaliteevski2006stability, lavrinenko2009influence, dorado2009modeling}. For example, only a few percent of disorder can eliminate the bandgap of inverse opal photonic crystals \cite{li2001photonic, li2000fragility}.  The only way to achieve robustness is to utilize nontrivial topological properties \cite{wang2009observation} in waveguides with gyromagnetic materials. However, the external magnetic field is a prerequisite to break time-reversal symmetry  \cite{lian2012robust}, hindering its practical application.

\begin{figure}[!t]
\includegraphics[width=\linewidth]{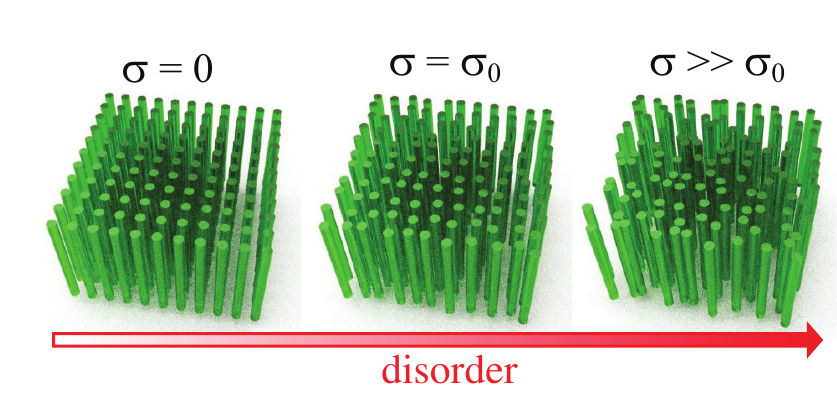}
\caption{
Schematic of a photonic structure, composed of dielectric nanorods, with an increasing position
disorder $\sigma$,  respectively. }
\label{fig:SchemeDisorder}
\end{figure}

\begin{figure*}[!t]
\includegraphics[width=0.8\linewidth]{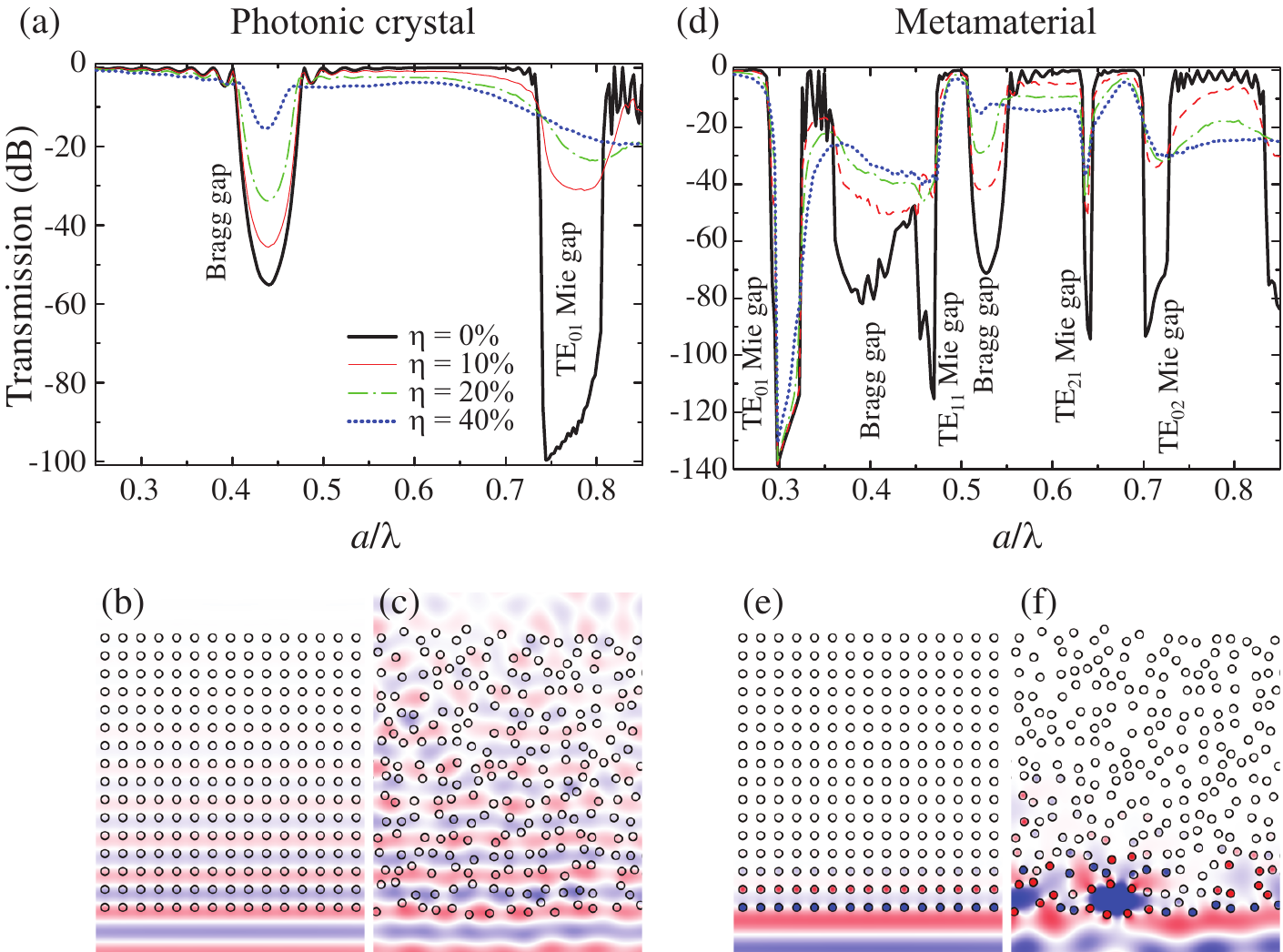}
\caption{ Transmission spectra of (a) a photonic crystal (with $\varepsilon=4$) and (d) a metamaterial (with $\varepsilon=25$) with different values of the disorder parameter $\eta$. Black solid curves are perfect structures ($\eta=0$), red dashed curves are weak disorder ($\eta=10\%$); green dash-dot curves are moderate disorder ($\eta=20\%$); blue dotted curves correspond to a strong disorder ($\eta=40\%$). Origin of spectral dips are labeled. Magnetic field distribution in (c,d) photonic crystal (for $\eta=0$ and $\eta=40\%$, respectively)  and (e,f) metamaterial (with $\eta=0$ and $\eta=40\%$, respectively). In each of the panels (c,d) and (e,f), the waves propagate from the bottom to the top.
}
\label{fig:Transmission}
\end{figure*}

Being inspired by the recent studies of the dielectric Mie-resonant metamaterials (MMs) and their link to PhCs~\cite{rybin2015phase, kruk2017functional, kivshar2018all}, here we consider photonic structures with the optically induced Mie resonances and reveal that they can support disorder-immune photonic bandgaps, in a sharp contrast with PhC where the Bragg resonances require stringent periodicity and consequently are not tolerant to disorder. Our numerical results demonstrate robustness of the optical waveguides under intense disorder, suggesting the way towards a new generation of disorder-immune photonic devices with cost-effective fabrication processes.

We start from an ideal periodic structure composed of nanorods arranged in a square lattice, as illustrated in the left panel of Fig.~\ref{fig:SchemeDisorder}. The lattice constant is $a=500$~nm and the rod radius is $r=125$ nm, so that the ratio $r/a$ defines a filling fraction of the structure. The permittivity of the nanorod is $\varepsilon$, with whose value identifying the system either as photonic crystals (low $\varepsilon$) or dielectric metamaterials (high $\varepsilon$)~\cite{rybin2015phase}. As the first step, we introduce disorder to the rod position $(x^i,y^i)$ as:  $x^i = x^i_0 + \sigma {U_x}$ and $y^i = y^i_0  +  \sigma {U_y}$, where $(x^i_0,y^i_0)$ is the original position in the periodic lattice, $U_x$ and $U_y$ are random variables distributed uniformly over the interval $[-1, 1]$, and the parameter $\sigma$ describes the strength of the disorder. We also consider the normalized disorder strength $\eta$ defined as the $\sigma$-to-$a$ ratio expressed in \%. Since the height of the nanorod is much larger than its radius, a two-dimensional approximation in the $(x, y)$ plane is valid.

When disorder is introduced (Fig.~\ref{fig:SchemeDisorder}), the translational symmetry of the structure becomes broken, making the bandgap structure of the spectrum in the reciprocal space to be ill-defined. As a result, we study properties of these photonic structures in the real space assuming that the low-transmission spectral regions associated with bandgaps are still observable in the corresponding spectrum.

In the analysis of disordered media, the light propagation is characterized by a logarithmic-average transmission instead of average transmission (see Ref.~\cite{Lifshits1988} for details). Figure~\ref{fig:Transmission} demonstrates the logarithmic-averaged transmission (averaged over an ensemble of 100 samples) vs. the disorder strength $\eta$ for both photonic crystals and metamaterials. In addition, we show the results for the wave propagation through the corresponding structures with a specific disorder realization for the lowest bandgap. We observe that the spectra consist of a number of pronounced dips (associated with the spectral gaps) which can be linked to either Mie and Bragg resonances. The Bragg gaps are observed as symmetric dips, while the Mie gaps have a knife-tip shape. In the regime of photonic crystals [Fig.~\ref{fig:Transmission}(a)],  we observe a degradation of all gaps with a stronger effect manifested for the second bandgap associated with  the TE$_{01}$ Mie resonances (about 70 dB for even weak disorder $\sigma=50$~nm or $\eta=10\%$).

In a sharp contrast, in the regime of a metamaterial [Fig.~\ref{fig:Transmission}(d)], the lowest bandgap survives under even strong disorder of $\sigma=200$~nm (or $\eta=40\%$).  In this regime, the Mie scattering from individual nanorods play a paramount role to form the bandgap through the TE$_{01}$ Mie resonances, reducing strict requirements of periodicity. The field distributions shown in Fig.~\ref{fig:Transmission}(e,f) reveal the effective field suppression by each nanorod oscillating out-of-phase with the incident wave.
Rigorous model accounts perturbations of coupling constants between neighbor rods \cite{dmitriev2019combining} due to the position disorder. It results in degradation of the suppression however this affects the TE$_{01}$ Mie gap much weaker than its Bragg counterpart.
Remarkably, the position disorder affects all other gaps including higher-order Mie gaps.
The reason is that Bragg frequency obey the law $f\propto(d \cos\theta)^{-1}$, where $d$ is a lattice spacing and $\theta$ is the propagation angle. The higher-order Mie gaps above the lowest Bragg gap do not demonstrate robustness, since they are not pure Mie gaps but mixtures with Bragg waves for certain directions. Thus, in spite of the identical configurations of the dielectric nanorods in Fig.~\ref{fig:Transmission}(c) and Fig.~\ref{fig:Transmission}(f), the wave propagation is remarkably different when the dielectric constant $\varepsilon$ of each rod changes from low to higher values.

\begin{figure}[!t]
\includegraphics[width=\linewidth]{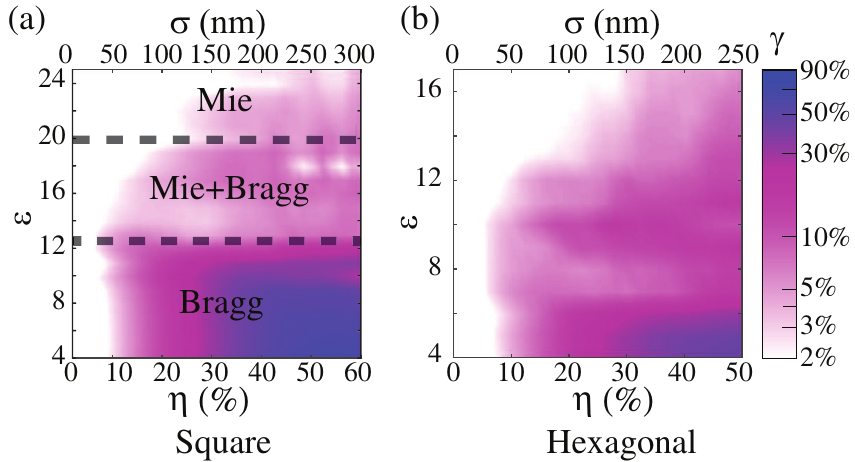}
\caption{
Degradation of the reflection $\gamma$
at different values of $\varepsilon$, illustrating
the robustness of the structure for (a) square and (b) hexagonal lattices, respectively.  }
\label{fig:DegradationMap}
\end{figure}

To provide a comprehensive picture of the impact of the position disorder to the system transforming from the PhC to MM regimes, we conduct a series of numerical simulations with different values of $\varepsilon$ ranging from 4 to 25. To quantitatively represent the robustness of the photonic bandgap, we define parameter
$\gamma = ({\overline{R}_0-\overline{R}})/{\overline{R}_0},$
where $\overline{R} = \int^{\lambda_2}_{\lambda_1} R d\lambda/(\lambda_2-\lambda_1)$ is the averaged reflection in the bandgap between $\lambda_1$  and $\lambda_2$; $\lambda_1$ and $\lambda_2$ are wavelengths at 10\% of the transmission minimum.

The degradation is normalized by $\overline{R}_0$, that is the reflection in the bandgap without disorder.
Consequently, $\gamma$ represents the deterioration of reflection in the bandgap, with a smaller value demonstrating a better robustness.
Figure~\ref{fig:DegradationMap}(a) summarizes the value of $\gamma$ with different $\varepsilon$ as the disorder $\sigma$ increases for the bandgaps with lowest energy (as shown in Figs.~\ref{fig:Transmission}(a) and (d). The value of $\gamma$ is averaged from three different sets of the uniform random variables. The relative change to the lattice constant is also labeled by the $\eta$ axis. The diagram can be unambiguously divided into three regimes. When the value of $\varepsilon$ is small, the system operates as a PhC, corresponding to the situation shown in Figs.~\ref{fig:Transmission}(a-c). The bandgap is quite vulnerable to the disorder, around 10\% of the position disorder can break the perfection of the photonic bandgap. As the increment of the permittivity, the system transforms into a new regime, where the bandgaps is formed by the overlap of Mie and Bragg resonances~\cite{rybin2015phase}. The Mie scattering from individual nanorods increases the robustness to the disorder, reducing the degradation $\gamma$ compared to the previous regime. Further enhancement of $\varepsilon$ drives the lowest bandgap formed by the Mie scattering, and the system transforms into the effective MM structure with a good robustness to the position disorder. Under intense disorder, a well-defined stop band persists, as illustrated in Figs.~\ref{fig:Transmission}(d-f). The transition of the robustness parameter $\gamma$ precisely matches the phase transition from PhC to MM "phases"~\cite{rybin2015phase}, identifying the unique role of the Mie scattering playing in the disorder-immune photonic bandgaps.

To illustrate generality of the disorder-immune photonic bandgaps, we analyze the structure robustness for a different geometry, namely a hexagonal lattice of nanorods that can support a bandgap for the TE waves~\cite{johnson2003introduction}. In addition, a different ratio $r/a=0.3$ is employed for the generality study while the lattice constant is kept the same, $a=500$~nm. Figure~\ref{fig:DegradationMap}(b) shows the value of $\gamma$ for varying $\varepsilon$. A similar behavior is illustrated for a square lattice nanorods in Fig.~\ref{fig:DegradationMap}(a), demonstrating three regimes with different level of robustness to the position disorder. The values of $\varepsilon$ for achieving strong robustness is ameliorated to a smaller value around 15 due to the optimal lattice \cite{li2018towards}. The typical transmission spectra for both PhC and MM regimes can be found in Supplemental Material \cite{SupplMater}.

\begin{figure}[!t]
\includegraphics[width=0.99\linewidth]{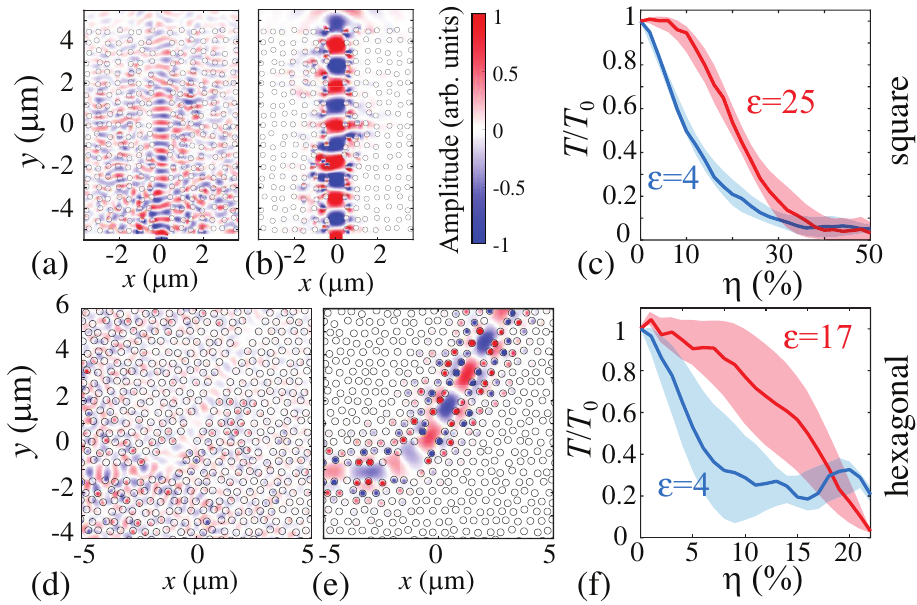}
\caption{(a,b) Field distribution $H_z$ in a straight waveguide created in a square lattice of nanorods with disorder $\eta=10$\% ($\sigma$ = 50~nm) for (a) $\varepsilon=4$ and (b) $\varepsilon=25$. (c) The corresponding relative transmission $T/T_0$ vs. $\eta$ for $\varepsilon=4$ and $\varepsilon=25$, respectively, for a square lattice of nanorods. (d,e) Field distribution $H_z$ in a bent waveguide created in a hexagonal lattice with disorder $\eta=8$\% ($\sigma$ = 40~nm) for (d) $\varepsilon=4$ and (e) $\varepsilon=17$. (f) The corresponding relative transmission $T/T_0$ vs. $\eta$ for $\varepsilon=4$ and $\varepsilon=17$, respectively, for a hexagonal lattice of nanorods.
}
\label{fig:WaveGuide}
\end{figure}

We further investigate the robustness effect for more practical structures such as a photonic waveguide. The waveguide is readily generated by introducing a line defect along the $y$-direction. For the PhC case, we select the TE$_{01}$ Mie bandgap with a better light confinement (since the Bragg frequency has a strong angular dependence). Figures~\ref{fig:WaveGuide}(a,b) illustrate the spatial distribution of the magnetic field in a disorder-impacted waveguide operating as PhC and MM, respectively, while the disorder-free example can be found in SM \cite{SupplMater}.
Here the disorder parameter $\sigma$=50~nm. In spite of the disorder destroying the ideal structure, the light is still well confined in the active region [Fig.~\ref{fig:WaveGuide}(b)], demonstrating a good robustness when operating in the MM regime.
With a reduced value of $\varepsilon$, the waveguide losses its function with the transverse diffusion under the same position configuration for a PhC, as shown in Fig.~\ref{fig:WaveGuide}a. To quantitatively demonstrate the impact of disorder, we calculated the transmission $T$ of the waveguide under different $\sigma$, as shown in Fig.~\ref{fig:WaveGuide}(c). Similarly, the transmission is normalized to $T_0$, the value without disorder for the comparison. A definite improvement of the robustness for the waveguide is observed for the MM regime with large permittivity. In addition, a more complicated situation is investigated with a bent waveguide embedded into a hexagonal lattice, as shown in Figs.~\ref{fig:WaveGuide}(d,e). Despite the vulnerability to the disorder at the corner where the propagation direction varies, the waveguide operating in the MM regime [Fig.~\ref{fig:WaveGuide}(e)] persists the function under the position disorder, compared with the PhC regime [Fig.~\ref{fig:WaveGuide}(d)].  We implement a quantitative analysis for the transmission degradation in Fig.~\ref{fig:WaveGuide}(f), demonstrating the robustness enhancement from permittivity increment similar to a straight waveguide in a square lattice of nanorods shown in Fig.~\ref{fig:WaveGuide}(c).

\begin{figure}[!t]
\includegraphics[width=\linewidth]{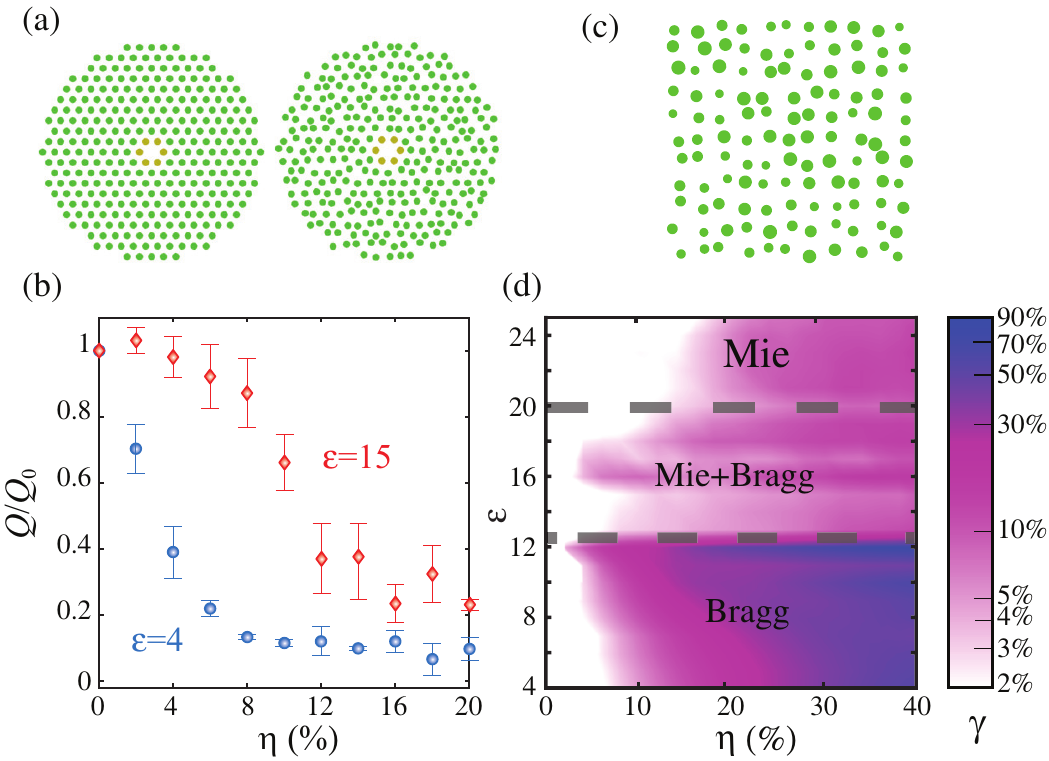}
\caption{
(a) An optical cavity formed by a hexagonal lattice of dielectric rods without disorder (left) and with disorder (right).
(b) Degradation of the relative quality factor $Q/Q_0$ with disorder $\sigma$ for $\varepsilon=4$ and $\varepsilon=15$.
(c) An example of a photonic structure with both position and size disorder, for $\eta=25\%$.
(d) Robustness parameter $\gamma$ vs. disorder $\eta$ and $\varepsilon$.
}
\label{fig:Qfactor}
\end{figure}

Besides waveguides, PhCs are widely used as to build optical cavities for different applications ranging from lasing to sensing.
Consequently, it is crucial to clarify the robustness of an optical cavity formed by high-index dielectric rods.
Here we create an optical cavity by introducing a point defect in the hexagonal lattice, as shown in Fig.~\ref{fig:Qfactor}(a).
Again, we assume that disorder is embedded in the rod position, as demonstrated in Fig.~\ref{fig:Qfactor}(a) for a perfect cavity ($\sigma=0$, upper panel) and a disordered cavity  ($\sigma=100$ nm, lower panel).

The position fluctuations for nanorods defining the cavity boundary (circles) is intentionally eliminated to provide a fixed cavity shape.
For the evaluation of the cavity robustness, we analyze the $Q$ factor calculated as $Q=\omega_R/\Delta\omega$ through the signals from four randomly located points inside the cavity, with $\omega_R$ being the resonant frequency, and $\Delta\omega$ being FWHM parameter.
Figure~\ref{fig:Qfactor}(b) demonstrates a relationship between $Q$ and $\sigma$ for both PhC and MM regimes, respectively, with the value averaged with three different sets of random variables. The $Q$ factor is normalized to its value without disorder, $Q_0$, to exclude a difference between the two cases. We observe that PhC structures ($\varepsilon=4$) are quite fragile to the disorder, whereas the photonic cavity
operating in the MM regime ($\varepsilon=15$) is more robust, only experiencing severe degradation when $\sigma$ reaches 50~nm.
With the introduction of moderate degree of disorder, the quality factor is improved with the value $Q/Q_0$ beyond 1, matching earlier results \cite{yamilov2004highest}.

In addition to fluctuations in the rod positions, we study also the robustness to a size disorder and consider a photonic structure of different nanorods, see Figs.~\ref{fig:Qfactor}(c-d). We assume that the disorder introduces fluctuations in the rod radii, $r = r + \sigma_{r} U_r$, where $U_r$  are random variables distributed uniformly over the interval $[-1, 1]$, and the relative disorder is: $\eta = \sigma/a = \sigma_{r}/r$. Figure~\ref{fig:Qfactor}(c) shows a typical disordered structure with $\eta = 25\%$. Similarly, we use the parameter $\gamma$ to evaluate the structure robustness, as shown in Fig.~\ref{fig:Qfactor}(d) for the regimes transform from PhC ($\varepsilon=4$) to MM ($\varepsilon=25$). The fluctuations in the nanorod radius cause a variation of the Mie resonance, consequently inducing deterioration of the robustness compared with the case presented in Fig.~\ref{fig:DegradationMap}(a). In additional, the same fluctuation in size $\sigma_r$ could bring stronger disorder to the nanorod with higher permittivity, considering the optical wavelength inside the dielectric $\lambda=\lambda_0/\sqrt{\varepsilon}$ and consequently increasing the ratio between $\sigma_r/\lambda$. This effect causes the robustness decreases as the increment of $\varepsilon$ in some regions. However, an obvious improvement in robustness is observed when the system works in the Mie regime as a metamaterial compared to photonic crystals in Bragg regime.

In summary, we have revealed a novel regime for the scattering of light in photonic structures with robust bandgaps by transforming
the structure from a photonic crystal to a dielectric metamaterial. When the Mie scattering from individual dielectric elements dominate over the Bragg scattering, both reflection and confinement of light becomes immune to an intense disorder.  Our study provides an useful guide for the nanofabrication of different photonic structures by employing dielectric metamaterials with high $\varepsilon$ for achieving the robust bandgap regime and also lifting strict requirements on periodicity. For hexagonal lattices, one can achieve robust bandgaps from the visible to infrared spectra for GaAs~\cite{aspnes1986optical} and Ge~\cite{jellison1992optical}. Importantly, such photonic structures can be realized with the bottom-up fabrication approach by utilizing the vertically aligned nanowires~\cite{persson2004solid, hersee2006controlled, wu2002growth}. In this case, the fluctuations in position dominates compared with that in size, as shown in Fig.\ref{fig:DegradationMap}(a).

\begin{acknowledgments}
The work has partially been supported by the ERC Consolidator Grant (TOPOLOGICAL), the Royal Society, Wolfson Foundation, the Ministry of Education and Science of the Russian Federation (3.1500.2017/4.6), the Russian Foundation for Basic Research (18-02-00427), the Australian Research Council, and the European Union Horizon 2020 research innovation programme under the Marie Sk\l{}odowska-Curie Grant (No. 752102). YK thanks K. Busch, Th. Krauss, and W. Vos for critical comments and useful discussions.
\end{acknowledgments}


\end{document}